\documentclass[journal]{IEEEtran}
\usepackage[utf8]{inputenc} 
\usepackage{comment}        
\usepackage[numbers]{natbib} 
\usepackage{hyperref}       
\usepackage{url}            
\usepackage{booktabs}       
\usepackage{amsfonts}       
\usepackage{nicefrac}       
\usepackage[activate=true]{microtype} 
\usepackage{lipsum}         
\usepackage{graphicx}       
\usepackage{array}          
\usepackage{doi}            
\usepackage{multicol}       
\usepackage{float}          
\usepackage{geometry} 
\usepackage{lipsum}   
\usepackage{xcolor}
\usepackage{tabularx} 
\usepackage{caption}  
\usepackage[justification=raggedright, singlelinecheck=false]{caption} 

\begin{document}

\title{Human-Computer Interaction and Human-AI Collaboration in Advanced Air Mobility: A Comprehensive Review}

\author{
    Fatma Yamac Sagirli, 
    Xiaopeng Zhao, 
    Zhenbo Wang 
    
    \thanks{Fatma Yamac Sagirli is a Ph.D. student in the Department of Mechanical, Aerospace, and Biomedical Engineering, University of Tennessee, Knoxville, TN 37996, USA (email: fyamacsa@vols.utk.edu).}

    \thanks{Xiaopeng Zhao is a Professor in the Department of Mechanical, Aerospace, and Biomedical Engineering, University of Tennessee, Knoxville, TN 37996, USA (email: xzhao9@utk.edu).}
    
    \thanks{Zhenbo Wang is an Associate Professor in the Department of Mechanical, Aerospace, and Biomedical Engineering, University of Tennessee, Knoxville, TN 37996, USA (email: zwang124@utk.edu).}
}

\markboth{Journal of \LaTeX\ Class Files,~Vol.~XX, No.~X, DECEMBER~20XX}%
{Yamac \MakeLowercase{\textit{et al.}}: Survey on Human-Computer Interaction, and Human-AI Collaboration in Advanced Air Mobility}

\maketitle

\begin{abstract}
The increasing rates of global urbanization and vehicle usage are leading to a shift of mobility to the third dimension—through Advanced Air Mobility (AAM)—offering a promising solution for faster, safer, cleaner, and more efficient transportation. As air transportation continues to evolve with more automated and autonomous systems, advancements in AAM require a deep understanding of human-computer interaction and human-AI collaboration to ensure safe and effective operations in complex urban and regional environments. 
There has been a significant increase in publications regarding these emerging applications; thus, there is a need to review developments in this area.
This paper comprehensively reviews the current state of research on human-computer interaction and human-AI collaboration in AAM. Specifically, we focus on AAM applications related to the design of human-machine interfaces for various uses, including pilot training, air traffic management, and the integration of AI-assisted decision-making systems with immersive technologies such as extended, virtual, mixed, and augmented reality devices.
Additionally, we provide a comprehensive analysis of the challenges AAM encounters in integrating human-computer frameworks, including unique challenges associated with these interactions, such as trust in AI systems and safety concerns. Finally, we highlight emerging opportunities and propose future research directions to bridge the gap between human factors and technological advancements in AAM.
\end{abstract}



\section{Introduction}
\vspace{12pt}

The emergence of AAM represents a significant evolution in transportation, driven by advancements in technology and the increasing demand for efficient urban and suburban mobility solutions. AAM encompasses a variety of applications, including air taxis \cite{rajendran2020air,lindner2024optimal}, cargo delivery \cite{gunady2022evaluating,german2018cargo}, and emergency medical services \cite{goyal2022advanced}, all of which rely on innovative aircraft designs, such as unmanned aerial vehicles (UAVs) and electric vertical takeoff and landing (eVTOL) vehicles. 
As cities become more congested, traditional ground transportation systems struggle to meet the demands of urban populations. AAM offers a potential solution to alleviate traffic congestion by utilizing airspace for short-distance travel, thus reducing travel times and enhancing overall mobility. This shift towards aerial transport is supported by advancements in autonomy, battery technology, digital communication systems, and intelligent technology, which collectively enable the operation of sophisticated autonomous vehicles capable of navigating above landscapes \citep{mathur2019paths,liu2023role}. 
However, the successful integration of AAM into existing transportation networks depends on several key factors, including user acceptance \citep{edwards2020evtol}, the development of regulations  \citep{faa2020urban}, and safe traffic management systems in high-density urban air spaces \citep{utm_conops_v2}.
Thus, the development of the AAM requires a deep understanding of human-computer interaction (HCI) to design interfaces capable of simulating real-world scenarios. 
\vspace{3pt}

The role of HCI in AAM is critical, as it is directly influenced by human factors like user experience and acceptance of these new technologies \citep{chauhan2021human}.
Effective HCI design must consider the unique challenges posed by aerial transport, including the need for real-time situational awareness \cite{kamkuimo2023decomposition}, intuitive control interfaces, and the management of user trust in autonomous systems.
Research indicates that trust is a pivotal factor in the acceptance of autonomous technologies, particularly in high-stakes environments such as AAM, where safety is paramount \citep{kim2023exploring}. 
Therefore, understanding how users interact with these systems and how they perceive the reliability of AI-driven decision-making processes is essential for fostering public confidence in AAM solutions. 
Moreover, the integration of AI into AAM systems presents both opportunities and challenges \citep{degas2022survey}. AI can enhance operational efficiency by optimizing flight paths, managing air traffic, and providing real-time data analytics for decision-making. However, the success of these AI systems depends on their ability to collaborate seamlessly with human operators. This collaboration requires a nuanced understanding of both human cognitive processes and AI capabilities, necessitating interdisciplinary research that bridges the gap between technical and theoretical components of human-AI interaction. 
\vspace{3pt}

While existing review papers focus on various technical aspects of AAM, such as the overall development of AAM systems \cite{tsai2023evtol,wang2023review,cohen2021urban}, eVTOL design and performance \cite{kiesewetter2023holistic,su2024evtol}, airspace design and management \cite{bauranov2021designing}, autonomous control and navigation strategies \cite{xiang2023autonomous,wei2024autonomous}, and the integration of artificial intelligence (AI) and explainable AI in air traffic management \cite{degas2022survey}, with particular attention to deep reinforcement learning approaches for conflict resolution \cite{wang2022review}, there is a noticeable gap in addressing the human-centric dimensions of these advancements. This paper aims to address this gap through three primary contributions:

\begin{itemize}
   
    \item \textbf{Focus on Human-Computer Interaction (HCI) and Interface Design:} By centering the review on HCI and interface design, the study highlights the importance of designing user-centric systems for effective interaction within AAM applications.
    
    \item \textbf{Exploration of Human-AI Collaboration:} The study investigates the development of collaborative algorithms that enhance the partnership between human operators and AI systems, ensuring seamless and efficient interaction in the AAM domain.
    
    \item \textbf{Examination of Trust in Autonomous Systems:} The paper analyzes mechanisms to foster trust in autonomous and AI-driven systems for future AAM applications, emphasizing their importance for user acceptance and system reliability.
    
\end{itemize}
\vspace{6pt}

The structure of this paper is organized as follows: Section~\ref{sec:methods} provides an overview of the AAM concept and its relationship to HCI. Section~\ref{sec:app} offers a detailed examination of human-computer interface designs, human-AI collaboration, and their applications in AAM. Section~\ref{sec:sec4} addresses the challenges related to HCI and AAM, offering insights into future trends. Finally, Section~\ref{sec:sec5} presents some concluding remarks.
\vspace{3pt}


\section{Methods and Theoretical Foundations}
\label{sec:methods}
\vspace{12pt}

\subsection{Review Method}
\vspace{3pt}

To ensure a comprehensive and unbiased review of current research on HCI and Human-AI Collaboration in AAM, we followed a systematic literature selection process. 
\vspace{3pt}

\textbf{Databases Searched:} We conducted our literature search using the following databases, which were selected for their relevance and comprehensive coverage of research in robotics, AI, and air transportation systems: IEEE Xplore, ACM Digital
Library, Scopus, Google Scholar, ArXiv, and conferences.
\vspace{3pt}

\textbf{Keywords and Search Strings:} We employed a combination of keywords and Boolean operators to identify relevant studies. 
We used following search terms were used: 
”Advanced Air Mobility” AND (”Human-Robot Interaction” OR ”Human-Computer Interaction” OR ”Human-Computer Interfaces” OR ”Human-AI Collaboration”).


\subsection{Definition and Scope of Advanced Air Mobility}
\vspace{6pt}

\begin{figure*}[!t] 
    \includegraphics[width=\textwidth]{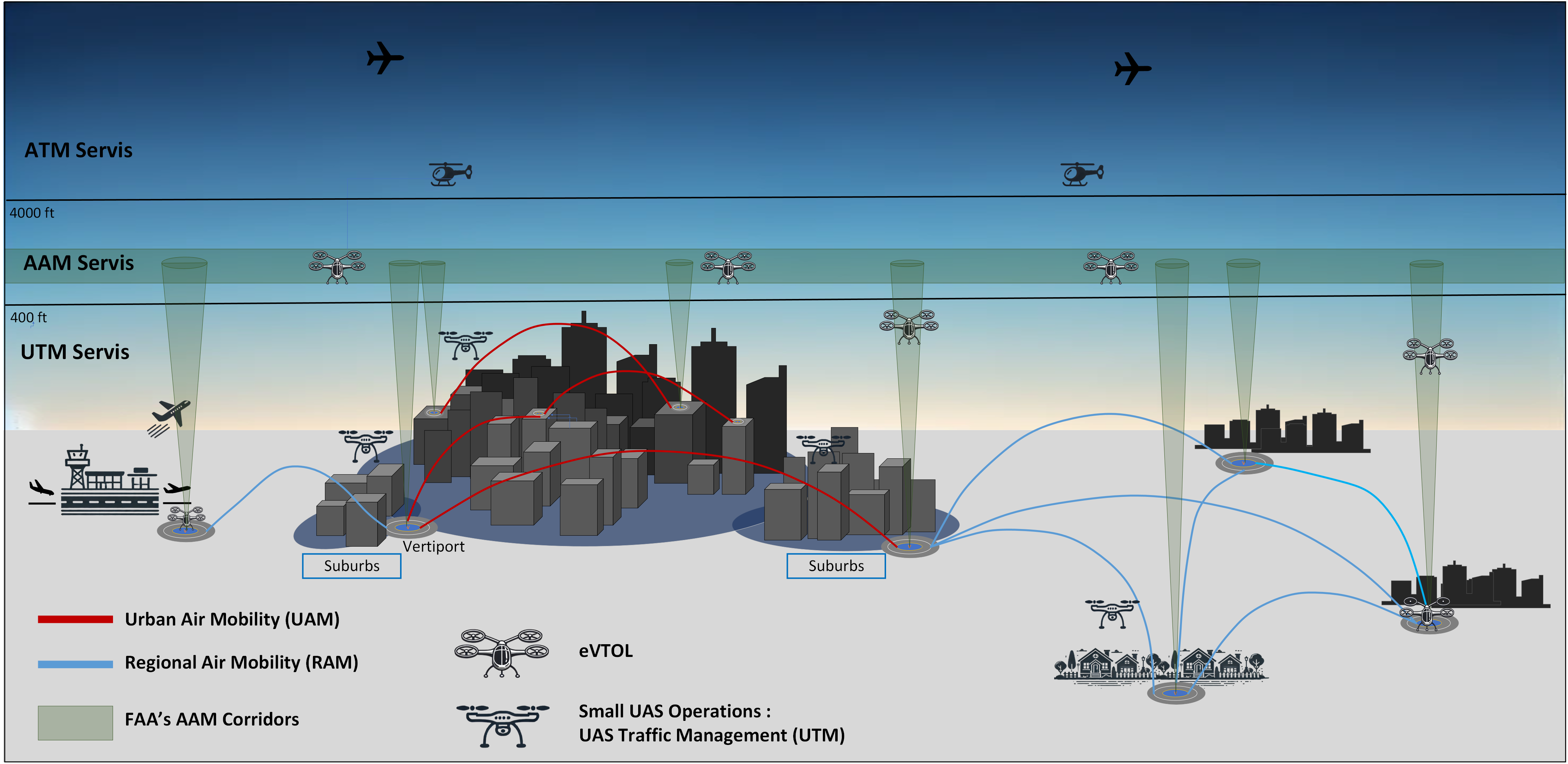} 
    \caption{Concept of AAM, UAM and RAM (Redrawn based on\citep{faa2020urban}) }
    \label{fig:aam}
\end{figure*}

AAM refers to a transformative approach to air transportation, driven by increasing congestion in ground transportation infrastructure due to population growth \cite{wild2024urban}.
It is a specialized concept developed by the Federal Aviation Administration (FAA), the National Aeronautics and Space Administration (NASA), and the aerospace industry \citep{faa2020urban}. 
Conventional air mobility relies on fixed-wing aircraft for general aviation, as well as short- and long-haul flights. It is primarily oriented towards connecting larger cities, countries, and continents, focusing on high-capacity and long-range travel.
AAM concept, depicted in Fig.~\ref{fig:aam}, is distinguished from conventional air mobility by its focus on urban and regional applications, aiming to enhance connectivity within cities and between urban centers and regional areas. 
Within the AAM framework, Urban Air Mobility (UAM) focuses on short-distance travel within inner-city areas, including connections between suburbs and city centers. It aims to optimize local transportation networks by reducing congestion and improving accessibility. In contrast, Regional Air Mobility (RAM) targets medium-distance transportation, connecting cities and remote areas to urban hubs. These capabilities collectively fill the gap between ground transportation and traditional aviation. 
\vspace{3pt}

AAM aims to alleviate ground traffic congestion and reduce commuter travel times for safe, low-carbon, and convenient transportation \citep{tsai2023evtol}. 
Recent improvements in electric propulsion and battery capacity have led to the development of eVTOLs aircrafts \cite{xiang2023autonomous}. 
These technologies enable more efficient and sustainable transportation methods while providing flexibility to access hard-to-reach areas. These advancements are not merely enhancements but are fundamental shifts designed to address increasing demands for safety, efficiency, and environmental sustainability in air travel. As AAM grows, integrating these technologies will redefine logistics, passenger transport, and emergency services. For instance, last-mile delivery services leverage UAVs to transport goods quickly, bypassing ground traffic and ensuring timely deliveries. Additionally, scheduled public air transportation offers a streamlined alternative for commuters, connecting key urban hubs with minimal delay. Furthermore, emergency air response services offer a transformative solution for metropolitan areas. They enable rapid, flexible access to critical care and assistance when every second counts, bypassing the limitations of traditional road transportation. 

\vspace{6pt}
\subsubsection{AAM Corridors}

The AAM corridor concept, illustrated in Fig.~\ref{fig:aam}, developed by the FAA, provides defined airspace pathways for alternative modes of transportation utilizing eVTOLs \citep{faa2020urban}. 
These corridors are part of a broader effort to integrate AAM into the National Airspace System (NAS) in a safe, efficient, and scalable manner, ensuring compatibility with existing air traffic. 
AAM corridors are reserved for low-altitude operations, typically below 4,000 feet, and are segregated from manned aviation to minimize the risk of mid-air collisions. 
These corridors will have pre-defined routes connecting key areas, such as urban centers, suburbs, and vertiports. 
Initially, the corridors will link two AAM vertiports to support direct operations. In later stages, the FAA anticipates the development of more complex and efficient networks that move beyond direct operations. 
The corridors act as a separation mechanism between AAM and other operations. Within these corridors, AAM operators are responsible for maintaining safe separation. During the initial phases of AAM operations, this requires having a pilot on board. However, in future operations, autonomous pilots may be an option.

\vspace{6pt}
\subsubsection{UAS Traffic Management (UTM)}

The complexity of air traffic is expected to rise significantly as the use of small Unmanned Aircraft Systems (UAS) expands and diversifies \cite{huang2024potential}. These operations are being employed for various applications, including package delivery, news collection, precision agriculture, infrastructure inspection, and disaster response. 
Consequently, ensuring the safety and security of frequent low-altitude urban flights is crucial to prevent potential hazards for people on the ground. Therefore, managing UAS traffic is a complex challenge that requires a combination of technologies, regulations, and operational strategies. 
UAS traffic management (UTM) project, developed by FAA and NASA, would integrate UAS operations in the airspace above buildings and below traditional aviation operations 
\cite{utm_conops_v2}. UTM operations take place in airspace below 400 feet, and their development is classified into four Technical Capability Levels (TCL) by the FAA based on population and traffic density: remote, sparse, moderate, and dense.


\subsection{Human-Computer Interaction in AAM}
\vspace{6pt}

\begin{figure*}[!t] 
   \includegraphics[width=\textwidth]{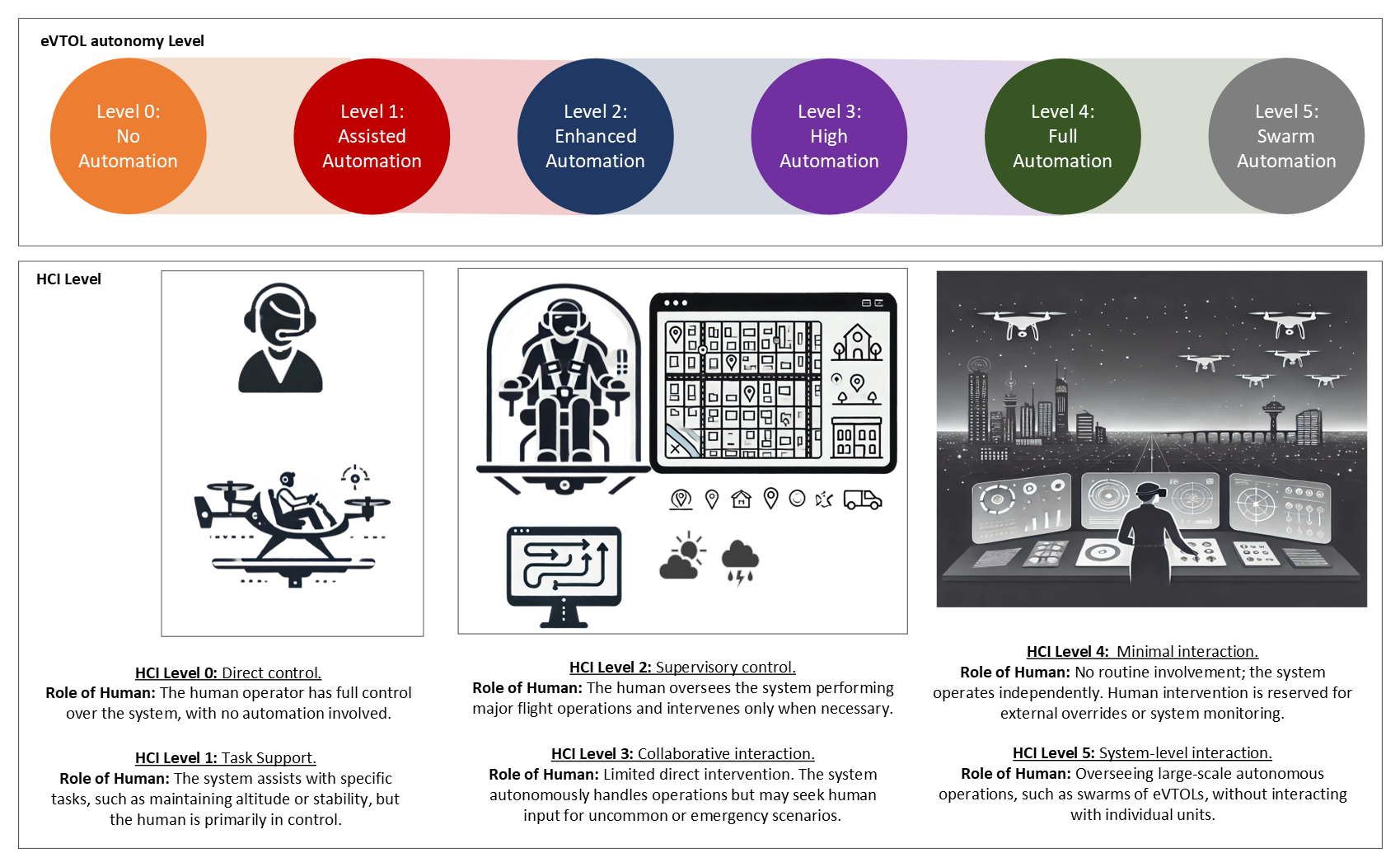} 
    \caption{ eVTOL autonomy and HCI Integration Levels for AAM (Redrawn based on \cite{wei2024autonomous}) }
    \label{fig:hci_levels}
\end{figure*}

Initial commercial AAM operations are expected to focus on delivering goods and transporting passengers. During the initial phase, a pilot will be onboard to ensure safety and control during passenger transportation, helping to provide a sense of security and comfort. 
However, as technology progresses, remote piloting and even fully autonomous services may become available, allowing passengers to travel more conveniently and efficiently while maintaining high safety standards.
Therefore, design safety and ensuring operational safety are crucial for determining how humans will engage with progressively autonomous systems \cite{prinzel2022human}.
This transition relies on the autonomy levels in eVTOLs defined by \cite{wei2024autonomous} and the maturity levels of the AAM concept described by \citep{goodrich2021description}.
To improve our understanding of the different levels of autonomy in AAM, we can incorporate HCI levels into this definition of autonomy. 
The Fig.~\ref{fig:hci_levels} illustrates the integration of HCI levels with the autonomy levels of eVTOLs and AAM, highlighting the evolving roles of humans across different stages of automation.
\vspace{3pt}

The transition emphasizes reducing human involvement while increasing the enhancement of automated systems. 
These levels progress from complete manual operation, where the pilot is solely responsible for all flight tasks, to full swarm automation, in which multiple eVTOLs operate collaboratively and autonomously without direct human intervention. Each level represents a unique combination of human involvement and automated capabilities, highlighting the potential for increased efficiency and safety in aerial mobility systems.
Progression towards higher levels of autonomy requires the effective design of HCI. As automation levels increase, the role of human operators evolves from active control to supervision and eventually to strategic oversight or no involvement. HCI connects humans and autonomous systems, ensuring that operations remain safe, efficient, and user-friendly across all levels of autonomy. The role of HCI in this development process can be described  below:

\vspace{6pt}
\subsubsection{Level 0–2: Supporting Pilots in Manual and Assisted Operations}

At lower autonomy levels, pilots keep significant control over the eVTOL. HCI design must focus on key factors such as intuitive cockpit interfaces with automation feedback systems. This interface provides pilots with an easy way to interpret flight information, such as navigation data, vehicle status, and environmental conditions, through user-friendly dashboards. Also, HCI should minimize the cognitive load by organizing information logically and using clear alerts for critical situations. In Levels 1 and 2, HCI must communicate what the automation is doing and where human intervention is necessary to prevent confusion or overreliance on incomplete automation.

\vspace{6pt}    
\subsubsection{Level 3–4: Supervisory Control and Trust in Automation}  
As the autonomy takes on more responsibilities, the pilot transitions to a supervisory role, which requires carefully designed HCI to maintain situational awareness and trust. In high level autonomy, automation systems must explain their decisions and actions to ensure operators understand why certain maneuvers are executed. Therefore, transparency and explainability are essential challenges in achieving a high automation level. 
Poorly designed systems may lead to over-reliance on automation, so HCI must ensure the operator's trust aligns with the system's decisions. 

\vspace{6pt}
\subsubsection{Level 5: Monitoring Swarm Automation}  

At the highest autonomy level, human involvement is minimal or limited to strategic oversight. Thus, HCI design might play a vital role in such applications:

\begin{itemize}
    \item Fleet Management Interfaces: Operators may supervise multiple eVTOLs or swarms in a UAM environment. The HCI should provide overal views of the fleet’s status, performance, and airspace coordination in a clear format.

     \item Human-Centered Automation: Even in highly autonomous systems, HCI should consider human oversight preferences, providing reassurance and control options when necessary.
    
    \item Decision Support Tools: In emergency scenarios, operators may need intuitive tools to interfere or adjust swarm behaviors quickly. These tools should be responsive to support fast decision-making.

\end{itemize}

Effective HCI design is fundamental to the success of AAM, enabling eVTOL operators to interact seamlessly with automation across all levels of autonomy. By prioritizing usability, safety, and transparency, HCI supports the transition from manual to autonomous operations, maintains situational awareness, and builds trust in automation. Without thoughtful HCI, the full potential of eVTOLs—particularly in high-density, urban environments—cannot be safely or efficiently realized.
\vspace{3pt}


\section{Applications}
\label{sec:app}
\vspace{12pt}

In this section, we will explore the applications of AAM in human-robot and human-computer interaction, as well as in interface design, which includes digital twins and mixed or virtual reality environments. Furthermore, we will examine studies on human-AI collaboration within the AAM framework. Our focus will be on how humans and AI systems can work together effectively to enhance operational efficiency and improve decision-making processes. Lastly, we will discuss important factors such as trust and acceptance in the adoption of these technologies. 
A summary of selected studies is presented in Table~\ref{tab:literature_applications}, organized by application cases and immersive technology used.


\subsection{Human-Machine Interface Designs}
\vspace{6pt}

In the concept of AAM, single-pilot operations are expected to be standard for several years before transitioning to remote operations or becoming fully autonomous.
Industry roadmaps \citep{european2020artificial} indicate that automation will progress from crew assistance between 2022 and 2025 to fully autonomous flights after 2035.
This shift will create a significant increase in the demand for pilots, which will require the development of training programs and enhanced cockpit automation to manage workload and ensure safety \cite{shi2023pilots, wechner2022design}. 
%
As AAM operations evolve, there is an increasing demand for comprehensive training and testing simulation platforms due to the complexities and challenges associated with high-density AAM operations in urban areas.
\vspace{3pt}

In recent years, the integration of immersive technology into AAM has gained significant attention due to its potential to enhance urban air space planning and operational efficiency for future AAM operations \cite{fakhraian2023towards,cross2022using}.
%
Namuduri \cite{namuduri2023digital} highlights the use of the digital twin approach for integrated airspace management to AAM and discusses how stakeholder communities - including academics, the Federal Aviation Administration, the aviation industry, and regional communities - are getting ready for this exciting development.
%
Moreover, game engine modeling and simulation using immersive technologies like Digital Twins (DT), Augmented Reality (AR), Mixed Reality (MR), and Virtual Reality (VR) offer potential solutions to this demand \cite{hulmegame,zintl2024mixed,turco2024study,ziakkas2024role,santhosh2023insights, zhao2024integrating}.
%
These simulators can create a safe, risk-free, and low-cost virtual environment where various flight scenarios can be tested, including different weather conditions \cite{tabassum2022preliminary}, fleet management for high-density environments \cite{hodell2022usability}, emergency events \cite{evtol_atc2023}, prototyping and testing of multi-agent solutions \cite{zhao2023digital,conrad2023developing}, collision detection and
situation awareness \cite{ywet2024yolotransfer}, demand-capacity balancing \cite{pongsakornsathien2021evolutionary}, along with many other purposes \cite{unverricht2022eye,reski2024designing}.
This integration can facilitate the training process while also prioritizing environmental sustainability by minimizing the carbon footprint associated with traditional training methods.
By effectively replicating real-world scenarios, these environments can help develop strategies to mitigate risks and improve the overall performance of AAM systems. 
Furthermore, effective training programs for future AAM pilots and instructors can be established to ensure that operators are well-experienced in the unique challenges and operational procedures of AAM vehicles \cite{schaffernak2022novel}. 
\vspace{3pt}

The study introduces DTUMOS \cite{yeon2023dtumos}, an open-source digital twin for urban mobility, combining AI-based estimated time of arrival models and vehicle routing for scalable and accurate operation. It supports iterative methods like reinforcement learning and provides a testbed for Mobility-as-a-Service (MaaS) and policy experimentation, linking MaaS with Digital-Twin-as-a-Service (DTaaS).
%
The framework proposed by Brunelli et al. \cite{brunelli2022framework} offers a novel approach to the development of urban aerial networks through the application of digital twin technology. Their study shows how digital twin technology can be applied through the example of Bologna city. It utilizes contextual data, including population figures, job locations, building types for obstacle clearance to identify suitable locations for vertiports and to determine the necessary urban aerial network. Their dynamic aerial network design enables real-time adjustments to activate or deactivate parts of the network based on traffic volumes, ensuring adequate separations and fast, competitive transport services. 
\vspace{3pt}

\begin{figure}[tbp] 
     \includegraphics[scale=0.46]{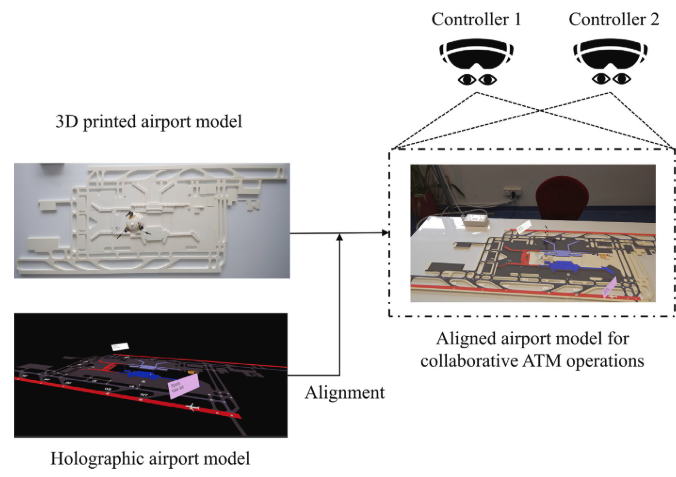}
    \caption{ Concept of the MR-based tangible airport digital tower system by \cite{chen2024tangible} }
    \label{fig:ref24}
\end{figure}

Another study on digital twin applications, conducted by Chen et al. \citep{chen2024tangible}, highlights the lack of research on visualization and interaction design of digital twins from a human factors perspective. Their research introduces a tangible digital twin framework that combines the 3D-printed Changi Airport model and its holographic 3D representation by projecting digital airport traffic over the 3D-printed version, illustrated in Fig.~\ref{fig:ref24}. The framework allows air traffic controllers (ATCOs) using MR headsets connected to the same network to perform ground control tasks. They can also interact with the system by touching the surface. To evaluate the overall usability of the system and the ATCOs' subjective perceptions, a mixed methods approach was utilized, incorporating both quantitative and qualitative measures. Three key human performance metrics were considered: perceived workload, situational awareness, and human trust. These metrics are essential when designing new human–machine interfaces for future air traffic management systems.
\vspace{3pt}

\begin{figure}[htbp] 
     \includegraphics[scale=0.37]{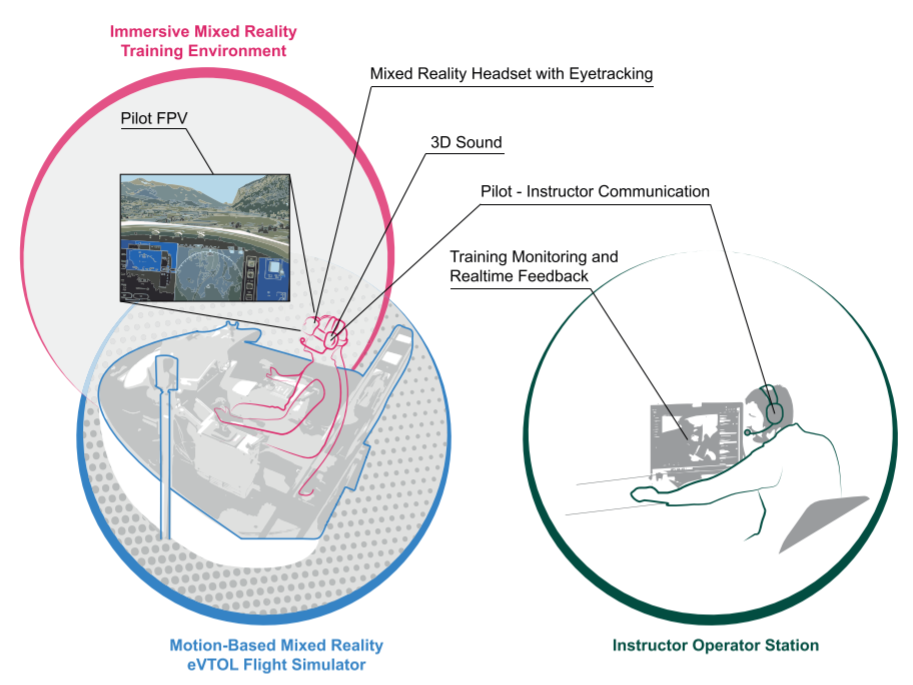} 
    \caption{Overview of an eVTOL MR simulator setup by \cite{kimura2024simulator} }
    \label{fig:ref26}
\end{figure}

A recent works \cite{zintl2022development,kimura2024simulator} on flight testing and pilot training for eVTOL aircraft introduce an innovative simulator-based MR approach to create immersive training environments that facilitate realistic flight simulations and improve pilot proficiency, illustrated in Fig.~\ref{fig:ref26}. Their study emphasizes the importance of integrating MR technologies to create immersive training environments that facilitate realistic flight simulations and improve pilot proficiency. This work not only contributes to the development of effective training methodologies for emerging aviation technologies but also highlights the potential for mixed reality to revolutionize pilot education in the context of urban air mobility.

\begin{figure}[htbp] 
     \includegraphics[scale=0.41]{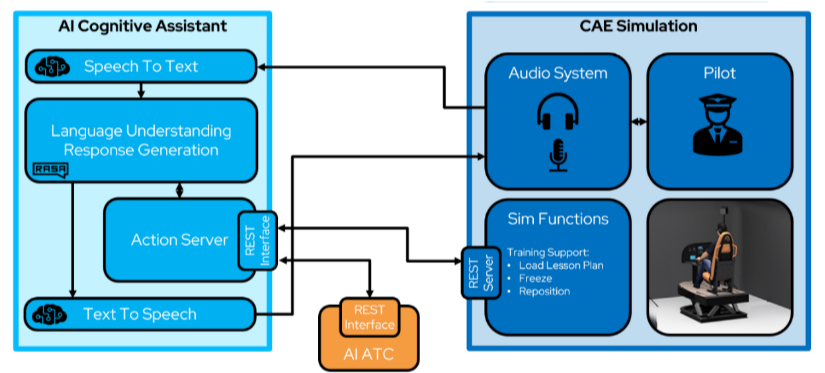} 
    \caption{Cognitive assistant architecture by \cite{evtol_atc2023} }
    \label{fig:ref27}
\end{figure}

%
Delisle et al. \cite{evtol_atc2023} explore the effectiveness of immersive mixed reality training for eVTOL emergency scenarios using an AI-powered Cognitive Agent to assist pilots, illustrated in Fig.~\ref{fig:ref27}. It evaluates AI performance metrics for a Natural Language Understanding (NLU) Dialog Manager and integrates the agent's dialogue into collaborative multi-agent reinforcement learning for air traffic control during emergency landings. The study also addresses robustness in reinforcement learning, evolutionary optimization, and nonlinear function approximation.

%
Apart from the AAM applications, the study conducted by Kim and Oh (2023) \citep{kim2024human} provides valuable insights into the advantages and challenges associated with human-embodied drone interfaces. They integrated immersive technologies (AR/VR) and haptics into the aerial manipulation. While their motivation lies in enhancing the abilities of workers in dangerous tasks using a mobile-manipulating unmanned aerial vehicle (MM-UAV), such designs may also be useful for future AAM cargo delivery tasks.


\subsection{Human-AI Collaboration}
\vspace{6pt}

As computer systems become more intelligent and autonomous, the functions of human operators are evolving. This shift is powered by integrating AI systems with human operators to enhance decision-making and operational efficiency. 
The integration of AI systems with human operators enhances decision-making and operational efficiency \cite{cocchioni2023learning,bhattacharyya2021assuring}. 
AI is commonly utilized in AAM, particularly with the growing trend of employing deep reinforcement learning algorithms for various applications. These include ensuring separation assurance \cite{brittain2021autonomous, deniz2024reinforcement}, balancing demand and capacity for conflict management \cite{chen2024integrated}, managing autonomous landings of eVTOL vehicles \cite{deniz2024autonomous}, and enhancing airspace structuring in UAM environment \cite{ribeiro2022using}. 
\vspace{3pt}

Given the expected growth of AAM operations in the coming decades, managing air traffic in urban airspaces will become increasingly complex, necessitating transparent AI support in UTM \cite{westin2020building}.
This complexity arises from the integration of a large number of UASs, including eVTOLs and drones, and the need for innovative solutions to ensure safe and efficient operations. 
As a result, there is a growing demand for advanced air traffic management approaches that can effectively handle the influx of air taxis, air ambulances, and air cargo transportation services in urban areas \citep{degas2022survey}. 
The concept of hybrid intelligence integrates human cognitive abilities with AI’s computational power, facilitating collaboration between humans and AI in future AAM operations \cite{chancey2023foundational}. For instance, the emerging field of human-in-the-loop AI and human-autonomy solutions for UTM \cite{lundberg2019human,lundberg2021human}, and human training \cite{byeon2024stochastic} highlights this integration. 
\vspace{3pt}

Krois et al. \cite{krois2024vertiport} highlight the need for vertiport infrastructure to support UAM and introduces the Vertiport Human Automation Teaming Toolbox, designed to facilitate real-time human-in-the-loop simulations for arrival, surface, and departure operations. 
Lim et al. \cite{lim2021adaptive} focuses on the development and evaluation of cognitive human-machine interfaces for supporting adaptive automation in unmanned aircraft system (UAS) operations. The study presents a system combining neurophysiological sensors and machine learning models to infer user cognitive states, with a specific scenario involving bushfire detection and UAV coordination. Through human-in-the-loop experiments, the system's ability to adapt in real-time was tested, showing potential for enhanced performance with further refinement of neurophysiological input features.
\vspace{3pt}

Pongsakornsathien et al. \cite{pongsakornsathien2020human} examine the need for an autonomous Decision Support System (DSS) for integrated manned/UAS Traffic Management (UTM) in urban airspace. The paper analyzes the roles and responsibilities of humans and systems to design effective Human-Machine Interfaces (HMI). The study highlights advanced traffic flow management concepts and introduces a Cognitive HMI concept to support closed-loop interactions and enhance system integrity.
Islam et al. \cite{islam2023wip} explore the dynamics of human-AI collaboration to emphasize the importance of effective HCI in AAM. The authors propose a reinforcement learning framework that facilitates adaptive interactions between humans and AI systems, allowing for improved decision-making and operational efficiency. Their findings indicate that well-designed AI systems can enhance situational awareness and support human operators in managing complex tasks compared to fully human or fully autonomous operations. The study highlights the need for ongoing research into the design of intuitive interfaces that can seamlessly integrate AI capabilities into human workflows. 
This research underscores the critical role of HCI in optimizing human-AI interactions in the rapidly evolving landscape of AAM.


\subsection{Trust, Acceptance and User Experience}
\vspace{6pt}

In the context of AAM, trust \cite{chancey2020designing,chancey2021enabling}, acceptance \cite{lotz2023user,kim2023exploring}, and user experience \cite{kim20211st} are interconnected factors that significantly influence the design of effective HCI. Trust is essential for the successful adoption of AAM systems, as it directly affects users' willingness to engage with autonomous technologies such as eVTOLs and drones. HCI design should enhance trust by emphasizing transparency, reliability, and feedback mechanisms. Transparent systems, for instance, clearly explain decisions and actions to users, such as why an autonomous eVTOL might alter its flight path or delay takeoff. 
\vspace{3pt}

Acceptance is closely related to user experience; if users find the interface intuitive and the interactions seamless, they are more likely to embrace the technology. 
As highlighted by Sato \cite{sato2023influence}, factors such as the perceived viability and independence of uncrewed air vehicles significantly influence public trust and acceptance of these technologies.
%
A recent study by Valente et al. \cite{valente2024towards} investigates the role of trust in HCI within the context of AAM. Their findings indicate that different flight phases—such as takeoff, cruising, and landing—along with varying visibility conditions (clear daylight, night, foggy), significantly affect passengers' trust levels in the system. The study reveals that adaptive augmented reality (AR) interfaces can enhance trust by providing real-time information and contextual awareness, which helps users feel more secure and informed during their journey. However, their AR interface is based on videos, and it is not an interactive flight simulation. 
\vspace{3pt}

Kim and Ji \cite{kim2024designing} investigate the impact of human-machine interface (HMI) design on passenger trust in eVTOL vehicles. Through an immersive simulator-based experiment involving 34 participants, the authors tested four distinct HMI conditions—baseline, movement, hazard detection, and full information—to assess their impact on passenger trust. The findings indicate that providing movement and hazard detection information significantly enhances passenger trust. Notably, the study also establishes a correlation between passengers' gaze behavior and their trust levels, suggesting that gaze patterns could serve as a valuable real-time indicator of trust. This pioneering research contributes essential insights for engineers and researchers focused on promoting the adoption of autonomous eVTOLs in urban environments, underscoring the necessity of optimizing HCI to enhance user experience and trust in these innovative transportation solutions.
Ultimately, a well-designed HCI framework that integrates trust, acceptance, and user experience can facilitate smoother interactions and promote the widespread adoption of AAM technologies, thereby enhancing their effectiveness and societal impact.

\begin{table*}[h!] 
\centering
\small  
\captionsetup{font=small} 
\caption{Summary of Selected Applications}
\begin{tabular}{@{}p{1.3cm}p{5.1cm}p{5.1cm}p{5.1cm}@{}}
\toprule
\textbf{References} & \textbf{Key Focus} & \textbf{Methodology} & \textbf{Findings}   \\
\midrule

\cite{brunelli2022framework}&Development of urban aerial networks using digital twin technology&Application of contextual data (population, job locations, building types)&Identified suitable vertiport locations and enabled real-time adjustments to aerial networks based on traffic volumes.\\

\citep{chen2024tangible}&Visualization and interaction design of digital twins &Tangible digital twin framework combining 3D-printed models and holographic representations&Enhanced air traffic controllers' (ATCOs) situational awareness and interaction through mixed reality headsets.\\

\cite{zintl2022development,kimura2024simulator}&Simulator-based MR approach for eVTOL pilot training&Immersive training environments for realistic flight simulations&Improved pilot proficiency and training methodologies for emerging aviation technologies.\\

\cite{evtol_atc2023}&MR training for eVTOL emergency scenarios&AI-powered Cognitive Agent assisting pilots&Evaluated AI performance metrics and integrated dialogue into multi-agent reinforcement learning for air traffic control.\\

\citep{kim2024human}&Human-embodied drone interfaces&Integration of AR/VR and haptics into aerial manipulation& Enhanced capabilities for workers in dangerous tasks, applicable to future AAM cargo delivery.\\

\cite{krois2024vertiport}&Vertiport infrastructure for UAM&Introduction of the Vertiport Human Automation Teaming Toolbox&Facilitated real-time human-in-the-loop simulations for arrival, surface, and departure operations.\\

\cite{lim2021adaptive}&Cognitive human-machine interfaces for UAS operations&Development of a system using neurophysiological sensors and machine learning &Tested real-time adaptation capabilities in a bushfire detection and UAV coordination scenario, showing potential for enhanced performance.\\

\cite{pongsakornsathien2020human}&Autonomous Decision Support System (DSS) for UTM&Analysis of human and system roles for effective Human-Machine Interfaces (HMI)&Introduced a Cognitive HMI concept to support closed-loop interactions and enhance system integrity in urban airspace.\\

\cite{islam2023wip}&Dynamics of human-AI collaboration in AAM &Reinforcement learning framework for adaptive interactions&Well-designed AI systems enhance situational awareness and support human operators in managing complex tasks compared to fully human or fully autonomous operations.\\

\cite{valente2024towards}&Trust in HCI within AAM& Investigation of trust levels across different flight phases and visibility conditions&Adaptive AR interfaces can enhance trust by providing real-time information and contextual awareness, although the interface is video-based and not interactive.\\

\cite{kim2024designing}& Impact of HMI design on passenger trust in eVTOL vehicles&Immersive simulator-based experiment with 34 participants testing four HMI conditions&Movement and hazard detection information significantly enhance passenger trust; gaze behavior correlates with trust levels, indicating potential for real-time trust indicators.\\

\bottomrule
\end{tabular}
\label{tab:literature_applications}
\end{table*} 


\section{Challenges and Future Directions}
\label{sec:sec4}
\vspace{12pt}

\subsection{Integration of Immersive Technology}
\vspace{6pt}

The integration of immersive technologies, such as extended, virtual, mixed, and augmented reality devices into AAM training and operations has the potential to enhance safety for both operators and passengers \cite{hurter2024past}. 
However, this integration also presents several challenges, including hardware limitations, software complexity, privacy concerns, and issues related to standardization \cite{lei2023virtual}.
Addressing these challenges requires a multifaceted research approach that encompasses technological development, usability testing, and the establishment of industry standards.
Hardware limitations, such as insufficient processing power, battery life, and display latency, hinder device portability and user comfort. On the software side, creating engaging and interactive AR/VR content is complex and resource-intensive, and designing intuitive user interfaces for immersive environments poses a significant challenge as users interact with 3D spaces in real time.
Additionally, privacy and security concerns, including data collection practices and exposure to cybersecurity threats, further complicate integration. 
Finally, interoperability with existing systems and the lack of standardization lead to fragmented implementations, limiting scalability and widespread adoption. Addressing these barriers requires collaboration among stakeholders to create efficient, user-friendly, and secure immersive ecosystems. 
While immersive technology has a long history, recent developments of immersive environments in AAM have lacked scientific interest. Future studies should examine the usability and effectiveness of these technologies in real-world AAM scenarios.


\subsection{Human-AI Collaboration}
\vspace{6pt}

Collaboration between humans and AI is becoming increasingly essential in both ground and air transportation, as the combination of AI systems with human operators can improve decision-making and operational efficiency.
The concept of hybrid intelligence, which combines human cognitive capabilities with AI's computational power, is particularly relevant in AAM domain.   
However, key issues include ensuring seamless human-AI interaction, such as smooth transition between AI and human control in autonomous vehicles, and addressing the variability in human behavior and expectations, which AI systems may not always account for \cite{ziakkas2024challenges}. 
%
Another issue that arises from human-AI collaboration is explainability and transparency. Current AI systems often lack transparency, making it difficult for human operators to interpret and act on their recommendations during critical operations \cite{degas2022survey}.
Data privacy and security also demand robust solutions to protect sensitive information while respecting user preferences. 
Lastly, societal acceptance remains a significant barrier, as public trust in AI technologies is limited due to concerns about safety. \cite{zanjanpour2024quantifying}. 
Despite recent advancements, these challenges remain unresolved in Human-AI Collaboration. Addressing these gaps is essential for ensuring the safety, efficiency, and scalability of AAM systems. Future research should focus on designing explainable AI systems capable of providing real-time, interpretable recommendations while balancing human control and autonomy.


\subsection{Airspace Design, Infrastructure Integration, and Environmental Constraints}
\vspace{6pt}

As AAM operations grow, the risk of congestion in urban airspace becomes a major concern. Wang et al. \cite{wang2021air} emphasize the need for air traffic assignment strategies that can effectively manage the increased volume of AAM operations while minimizing disruptions to existing air traffic. 
While the authors express significant concern regarding this issue, current research has predominantly prioritized theoretical concepts over empirical experimentation. Immersive environments, such as virtual or augmented reality simulations, present valuable opportunities to create safe and controlled testing grounds for exploring this problem. These environments allow researchers to conduct experiments that would be difficult or impossible to perform in real-world settings without risk. 
Future research initiatives could focus on the design of airspace and the evaluation of capacity by simulating various traffic scenarios within immersive environments. This approach could help researchers better understand the dynamics of air traffic management, identify potential issues, and develop innovative solutions to improve efficiency and safety in airspace operations.
\vspace{3pt}

Integrating AAM into the current airspace requires significant upgrades to existing infrastructure \cite{takacs2022infrastructural}. This includes the development of vertiports and high-capacity charging stations for vehicles. Urban planning is essential for determining the placement of vertiports and flight corridors. Factors such as land use, population density, and potential environmental impacts must be considered to avoid conflicts and ensure community acceptance \cite{cohen2024planning}. 
The absence of standardized vertiport designs hinders widespread AAM deployment. HCI can address this issue by offering virtual designs to plan the entire system prior to real-time implementation, allowing for the evaluation of the model's performance.
Future work should focus on developing modular HCI that offers a virtual template for vertiport designs, integrating seamlessly with existing infrastructure.
\vspace{3pt}

The operation of eVTOLs and drones in urban areas raises concerns about noise levels and their impact on communities \cite{jackson2023some}. While electric propulsion systems are generally quieter than traditional aircraft engines, the unique noise signatures of eVTOLs, particularly during takeoff and landing, can still be disruptive. 
Therefore, the environmental implications of AAM operations, particularly noise pollution, require further study. 
Future research should focus on designing quieter airspace for AAM operations through HCI interfaces to evaluate noise levels, thereby improving acceptance among urban populations.
%

\section{Conclusion}
\label{sec:sec5}
\vspace{12pt}

This review paper offers a brief summary of current research on HCI and human-AI collaboration in AAM. 
It specifically addresses the aspects of interface design using immersive technologies including VR and AR interfaces, along with human-AI collaboration for air traffic management in this emerging field. 
HCI and human-AI collaboration are indispensable for advancing AAM into a safe, efficient, and scalable transportation system. HCI bridges the gap between humans and complex autonomous systems, ensuring intuitive interfaces, situational awareness, and trust. Simultaneously, human-AI collaboration leverages the strengths of both humans and AI to optimize decision-making, manage air traffic, and enhance safety.
As AAM continues to evolve, the integration of well-designed HCI and robust human-AI collaboration will be key to realizing a future where autonomous aerial systems coexist seamlessly with human operators, passengers, and communities, enabling transformative mobility solutions for urban and rural spaces.

\ifCLASSOPTIONcaptionsoff
  \newpage
\fi

\bibliographystyle{unsrt}

\end{document}